%% file: Galli__Final_ICC18__-_Bare_Metal_Open_Waveguide_vfinal_-_ArXiv.tex
\documentclass[12pt,draftcls,onecolumn]{IEEEtran}
\usepackage[english]{babel}
\usepackage[T1]{fontenc} % uses 8-bit encoded glyphs
\usepackage{mathrsfs} % Ralph Smith's Formal Script Font, in addition to mathcal
\usepackage{verbatim}
\usepackage{comment}

\usepackage{latexsym, amssymb, dsfont, stfloats, color, url}
\usepackage[cmex10]{amsmath}  % Ensures that document uses only Type 1 fonts
\usepackage{graphicx}
\usepackage{epsfig}
\usepackage{float}
\usepackage[caption=false,font=footnotesize]{subfig} % Prevents automatic loading of Axel Sommerfeldt's caption.sty which will override IEEEtran.cls

\usepackage{array}    % Extending the array and tabular environments
\usepackage{tabulary} %  Similar to tabular environment, but you can set 'total width' and column specifications.
\usepackage{multirow} %  Create tabular cells spanning multiple rows

\usepackage{cite}
\usepackage{cases}    % Numbered cases environment
\usepackage{url}
\usepackage{balance}

\usepackage{nicefrac}

\DeclareGraphicsExtensions{.eps,.pdf,.png,.jpg,.gif,.jpeg,.pstex}

\input{MyMacros.tex}

\newcommand{\insertonefig}[5]
{
\begin{figure}[!htb]
\begin{center}
\includegraphics[angle=#1,width=#2,clip,keepaspectratio]{#3}
\end{center}
\caption{#5}
{#4}
\end{figure}
}

\begin{document}

% Paper title

\title{Bare Metal Wires as Open Waveguides, with Applications to 5G}

% For Drafts and Transactions
\author{
        Stefano~Galli,~\IEEEmembership{Fellow,~IEEE,}
        Ju~Liu,
        Guanxi~Zhang % <-this % stops a space

\thanks{S. Galli (stefano.galli@huawei.com) is with Futurewei Techncologies, Inc., Bridgewater, New Jersey, USA.}% <-this % stops a space

\thanks{Ju Liu and Guanxi Zhang are with Huawei Techncologies, Inc., Shanhai, China.}% <-this % stops a space

} % end of \author

% The paper headers - \MakeLowercase{...}
\markboth{Galli et al., ``The usual problem solved in the usual way''}%
{Submitted to the 2018 IEEE International Conference on Communications.}

% Make the title area
\maketitle

%\begin{singlespace}
\begin{abstract}
We study an old but unconventional waveguiding technique based on the use of overhead bare metal Medium Voltage (MV) power lines as \emph{open waveguides} and not as transmission lines. This technique can be used in support of 5G applications such as backhauling and broadband access in the mmW frequency range or above. Although the analysis of open waveguides dates back a century or more, very few communications oriented papers are available in the literature and there is today no commonly agreed upon channel model. This paper offers a first step in covering this gap by first solving numerically the characteristic equation derived from Maxwell equations, then deriving an analytical expression for the transfer function, and finally reporting for the first time capacity values for bare metal open waveguides. A channel capacity of $1$~Tbps over $100$~m can be achieved using a single overhead Medium Voltage power line, $1$~W of total transmitted power, and the $1-100$~GHz band.
\end{abstract}
%\end{singlespace}

\thispagestyle{empty}
%\newpage

%\IEEEpeerreviewmaketitle

\section{Introduction} \label{sec.Intro}
5G mobile wireless networks are expected to deliver a variety of non-incremental improvements over the previous generation. Although some wonder whether 5G is just a supply-chain induced bubble rather than a solution that consumers and verticals really need, there is no doubt that 5G technology is today the object of unprecedented interest and high expectations. As higher data rates and carrier frequencies will characterize 5G networks \cite{Andrews2014What5G}, cells will inevitably become smaller to increase area spectral efficiency and this will require easy access to backhaul capacity.

As cell density grows, backhaul availability needs may require the use of utility poles, towers, rooftops, down to street furniture and light poles.
Despite the availability of many wired and wireless technologies, the difficulty of providing ubiquitous high speed backhauling capable of supporting the low latency ($<150~\mu s$) requirements necessary for Centralized Radio Access Network (C-RAN) and Coordinated Multi-point Processing (CoMP) will likely be there for some time \cite{Jaber2016-5Gbackhaul}.

This paper describes an unconventional but promising backhauling technique which uses overhead bare metal Medium Voltage (MV) power lines as \textit{open waveguides} \cite{TR:Corridor:surface2009, Link:AirGig-2016}. We note that this way of using power lines for telecommunications is very different from conventional Power Line Communications (PLC) \cite{GalliScagWang2011ProcIEEE}, where power lines are used as \textit{transmission lines}. When using power lines as open waveguides, the excited (principal) mode actually propagates \textit{around} the power line as a Transverse Magnetic (TM) mode and not \textit{through} it as a Transverse Electromagnetic Mode (TEM).

The basic theory of using wires as open waveguides was developed over a century ago by Sommerfeld \cite{Sommerfeld1899}, extended in the 1950s by Goubau to the case of metal waveguides coated with a dielectric \cite{Goubau1950SWapplications,Goubau1951SC-SWTL}, and then nearly ``forgotten'' for $50$ years as, despite its appealing low attenuation, open waveguides found little or no practical applications. A primary reason for this is that experiments often could not detect the ``surface'' wave predicted by Sommerfeld. In fact, guided, partially guided, and radiated waves are generally simultaneously present, and it is the kind of excitation that determines which type of wave dominates -- see \cite{Dyott1952launching, Goubau1956OpenWires, Brown1959Annular, Elmore2012qex} for additional considerations and examples on launching devices. Another reason that made impractical Sommerfeld surface waves is their large field extension around the wire, which could be as large as several meters at sub-GHz frequencies. Thus, a bare metal open waveguide would require a lot of free space around it in order to let the wave propagate undisturbed, as wave propagation would be affected by bends and discontinuities along the line.

In the last decade, the topic of open waveguides was rediscovered. In 2004, Wang and Mittleman were the first ones to suggest that a simple bare metal wire could be used as a waveguide ``\textit{to transport terahertz pulses with virtually no dispersion, low attenuation, and with remarkable structural simplicity}'' and for applications in sensing, imaging and spectroscopy \cite{Wang2004nature}.

The first recent mention of Sommerfeld guided waves for telecommunications applications was made by Glenn Elmore of Corridor who proposed using overhead bare metal MV power lines as open waveguides to enable a ``\emph{Distributed Antenna System (DAS) network for wireless mobile service providers''} \cite{TR:Corridor:DAS2011} and also to solve the last mile broadband access problem \cite{TR:Corridor:surface2009}. In a 2016 press release, AT\&T disclosed the carrier's plans to use MV power lines as open waveguides for delivering ``\textit{ultra-fast wireless connectivity to any home or handheld wireless device}'' \cite{Link:AirGig-2016}. Finally, in  2017 Prof. Cioffi disclosed a novel method for providing Terabits per second over existing Digital Subscriber Lines (DSL) by exciting a guided mode between twisted-pairs within a DSL binder and using MIMO-like processing to handle the various modes propagating along the waveguide\cite{TR:ASSIA:TDSL2017}. Although propagation in TDSL is still guided, the propagating wave is likely a Surface Plasmon Polaritron which has a very different nature from the Sommerfeld  waves addressed here and it is excited at infrared frequencies or higher \cite{SarkarPalma2017}.

Despite this renewed interest, very few communications oriented papers are available in the open-waveguiding literature and a commonly agreed upon channel model is still lacking.

This paper provides a numerical solution to the characteristic equation of the bare metal open waveguide derived by Sommerfeld, bypassing the need of making frequency dependent approximations and making available the axial propagation constant $h$ over a very wide frequency range (from Hz to PHz). Then an analytical expression for the channel transfer function is derived and used to calculate important channel characteristics such as Root-Mean-Square Delay-Spread (RMS-DS), group velocity, and channel capacity. To the best of the authors' knowledge, these communications quantities are reported for the first time for bare metal open waveguides.

\section{Problem Formulation} \label{sec.ProblemForm}
%The first to investigate the propagation of electromagnetic waves along a metallic wire was Hertz in 1888. The launched waves exhibited a field that was loosely confined to the wire, so that reflection from walls complicated measurements. In 1890, Lecher introduced the two wire system which had much better field confinement and guiding properties than the single wire \cite{Lecher1890}. The Lecher two-wire system had good success and can be considered as the oldest open waveguide.
%
%In 1899, Sommerfeld solved the problem of a wave propagating along a bare metal, infinitely long, and straight wire with finite conductivity and immersed in air \cite{Sommerfeld1899}. The found solution was a non-radiating wave guided by and along the surface of the wire that exhibited an attenuation that was found to be much smaller than that of wireless free space propagation and waves in other transmission lines such as coaxial cables. Harms then extended Sommerfeld's results to the case of an insulated wire \cite{Harms1907}.

Let us consider the problem of a wave propagating on a circular cylinder of metal wire immersed in air. This is shown in Figure \ref{fig.CoordinateSystem}: the central region is a conductor of radius $a$ with permittivity $\epsilon_c$, permeability $\mu_c$, conductivity $\sigma_c$, and all is in air with permittivity $\epsilon_a$ and permittivity $\mu_a$.
Hereafter, subscripts ``0'', ``c'', and ``a'' in permittivity $\epsilon$, permeability $\mu$, conductivity $\sigma$, radial propagation constant $\lambda$, and wavenumber $k$ shall denote vacuum, conductor, and air. Finally, $\epsilon^{(medium)}_r,~\mu^{(medium)}_r$ shall denote the relative permittivity and permeability of a medium.
\insertonefig{0}{6.8cm}{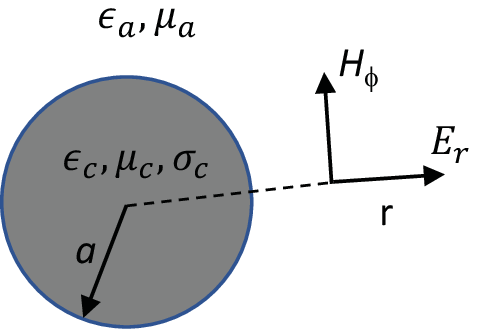}{\label{fig.CoordinateSystem}}
{Cross section of the guide, perpendicular to the propagation direction $z$. The direction of $E_z$ is perpendicular to this page.}

Using the coordinate system in Figure \ref{fig.CoordinateSystem}, the general solution of the fields inside conductor, dielectric, and air at a distance $r$ from the center of the conductor is given by (see \cite{Sommerfeld1899} and \cite{Stratton1941book}):
\begin{IEEEeqnarray}{rCl}
  E_r(z,r,f,t)     &=& A \frac{jh}{\lambda}Z_1(\lambda r) e^{j(2\pi f t - hz)}   \label{eq.CIL.Er} \\
  E_z(z,r,f,t)     &=& A  Z_0(\lambda r) e^{j(2\pi f t - hz)}  \label{eq.CIL.Ez} \\
  H_\phi(z,r,f,t)  &=& A \frac{j(\lambda^2+h^2)}{2\pi f \mu_r \lambda } Z_1(\lambda r) e^{j(2\pi f t - hz)}     \label{eq.CIL.Hphi}
\end{IEEEeqnarray}
where we have:
\begin{enumerate}
  \item $Z_n$ ($n=0, 1$) are the usual cylinder functions which can be expressed as either a linear combination of Bessel functions of the first ($J_n$) and second kind ($Y_n$), or of Hankel functions ($H_n=J_n\pm Y_n$).
  \item $\lambda$ is the \textit{radial} propagation constant, i.e. orthogonal to the direction $z$ of the metal cable.
  \item $h$ is the \emph{axial} propagation constant, i.e. along the direction $z$ of the metal cable, and it is the same in both conductor and air\footnote{In common engineering notation, the axial propagation constant is $\gamma=jh=\alpha+j\beta$, where  $\alpha=-\Im[h]$ and $\beta=\Re[h]$ are the attenuation (neper per meter) and phase (radians per meter) constants, respectively.}.
  \item $\mu_r$ is the relative permeability of the medium.
  \item $A$ is a constant determined by boundary conditions.
\end{enumerate}

The field components can be obtained by applying the boundary conditions below:
\begin{enumerate}
  \item The tangential component of $E_z$  and $H_\phi$ fields must be continuous at the surface of the conductor ($r=a$).
  \item Inside the conductor ($r\leq a$) fields must be finite, thus the solution is expressed in terms of Bessel functions.
  \item Outside the guide, i.e. in air ($r \geq a$) the field must decay to zero for $r\rightarrow\infty$, thus the solution must be expressed in terms of Hankel functions.
\end{enumerate}

For numerical computations, we will use the values shown in Table \ref{tab.NumValues}.

\begin{table}[htbp]
\centering
\caption{Values used in the simulations}
\label{tab.NumValues}

{

\begin{tabular}{|l|c|c|c|}
  \hline
            & $\epsilonv_\br$                          & $\muv_\br$    & $\sigmav_\bc$ \\
   \hline
   {Vacuum} & $1$                                   & $1$        & -   \\
   {Air}    &$\simeq 1$                            & $\simeq 1$ & -  \\
   {Copper} & $1-\frac{\sigma_c}{2\pi f \epsilon_0}$ & $\simeq 1$ & $5.96\cdot 10^7$ \\
%   {Polyethylene} & $2.25$                                 & $\simeq 1$ &  - & $3.1\cdot 10^{-4}$ \\
%   &  &  &  &  \\
  \hline
\end{tabular}
}
\end{table}

\section{Bare Metal Wire Case (Sommerfeld-wave)} \label{sec.Sommerfled}
A good overview of this case can be found in the book by Stratton \cite{Stratton1941book}. Stratton reports several findings on the propagating modes:
\begin{enumerate}
  \item There are an infinite number of symmetric and asymmetric electric and magnetic waves of propagation along a solid conducting cylinder each characterized by a discrete set of propagation constants $h_{n,m}$.
  \item If the conductivity of either medium is infinite, then the propagating waves must be either TM or TE.
  \item If the conductivity is finite, a superposition of electric and magnetic type waves is necessary to satisfy the boundary conditions, except for the symmetric ($n=0$) excitation.
  \item Among the infinite symmetric modes $h_{0,m}$, there is only one mode characterized by small attenuation and this \textit{principal} mode is a TM mode.
  \item All other modes (axially symmetric TE and asymmetric modes) are damped out quickly, within a few centimeters even at frequencies as low as 1 Hz.
  \item As the conductivity of the cylinder is decreased, the attenuation of the principal wave increases.
  \end{enumerate}
In this case, the cylinder functions describing the field inside the conductor are the bessel functions of the first and second kind as they are finite for $r=0$. For the field outside the conductor and in air, the Hankel functions of the first kind $H_0^{(1)}$ and $H_1^{(1)}$ are taken as they ensure the proper decrease of the field when $r \rightarrow \infty$.

At a distance $r$ from the center of the conductor, the cylinder functions take the following forms \cite{Stratton1941book}:
\begin{IEEEeqnarray}{rCl}
  Z_0(\lambda r) &=&
    \begin{cases}
        J_0(\lambda_c r)        & 0 \leq r \leq a  \label{eq.SW.Z0.in} \\
        H_0^{(1)}(\lambda_a r)  & a \leq r < \infty    \label{eq.SW.Z0.out}
    \end{cases}
\end{IEEEeqnarray}
\begin{IEEEeqnarray}{rCl}
  Z_1(\lambda r) &=&
    \begin{cases}
        J_1(\lambda_c r)         & 0 \leq r \leq a   \label{eq.SW.Z1.in} \\
        H_1^{(1)}(\lambda_a r)   & a \leq r < \infty \label{eq.SW.Z1.out}
    \end{cases}
\end{IEEEeqnarray}
where $\lambda_c$ and $\lambda_a$ are the radial propagation constants of the modes propagating inside the conductor and in air.

\subsection{The characteristic equation for the \textit{principal} TM mode }
The $E_z$ and $H_\phi$ fields must be continuous at the surface of the conductor. Equating the ratio of $E_z$ and $H_\phi$ at $r=a$ using the equations valid inside and outside the conductor, allows us to find the following characteristic equation for the principal TM mode \cite{Stratton1941book}:
\begin{equation}\label{eq.SW.TMequation}
  \frac{\lambda^2_c + h^2}{\mu_r^{(c)} \lambda_c} \frac{J_1(\lambda_c a)}{J_0(\lambda_c a)} = \frac{\lambda^2_a + h^2}{\mu_r^{(a)} \lambda_a} \frac{H^{(1)}_1(\lambda_a a)}{H^{(1)}_0(\lambda_a a)}.
\end{equation}
A variety of authors have solved \eqref{eq.SW.TMequation} making approximations appropriate for low frequencies \cite{Sommerfeld1899, Stratton1941book, Goubau1950SWapplications}. King and Wiltse solved \eqref{eq.SW.TMequation} in 1962 for the first time for frequencies in the THz region \cite{KingWiltse1962mmw}.

In this paper, rather than making approximations based on the frequency range of interest, we forgo the desire of an analytical solution, and we manipulate \eqref{eq.SW.TMequation} in a way that lends itself to being solved numerically via iterative methods. This method is applicable at any frequency range provided that a good frequency dependent starting point is used.

The first step is to express \eqref{eq.SW.TMequation} in implicit form. From the fact that $h$ must be the same for air and conductor, it descends that $\lambda_a^2-k_a^2=\lambda_c^2-k_c^2$. Substituting the former in \eqref{eq.SW.TMequation}, we obtain the following equation implicit in $\lambda_a$:
\begin{equation}\label{eq.SW.TMeqIMPLICIT}
  \lambda_a = \frac{\epsilon_r^{(a)}}{\epsilon_r^{(c)}} \frac{H^{(1)}_1(\lambda_a a)}{H^{(1)}_0(\lambda_a a)}  \frac{J_0(\lambda_c a)}{J_1(\lambda_c a)} \sqrt{k_0^2(\epsilon_r^{(c)}\mu_r^{(c)}-\epsilon_r^{(a)}\mu_r^{(c)})+\lambda^2_a}
\end{equation}
where:
\begin{itemize}
  \item $\lambda_a^2=k_a^2 - h^2$, with $k_a^2 = (2\pi f)^2 \mu_a \epsilon_a = k^2_0 \mu_r^{(a)} \epsilon_r^{(a)}$.
  \item $\lambda_c^2=k_c^2 - h^2$, with $k_c^2 = k^2_0 \mu_r^{(c)} \epsilon_r^{(c)}$.
%  \item $k_a^2 = (2\pi f)^2 \mu_a \epsilon_a = k^2_0 \mu_r^{(a)} \epsilon_r^{(a)}$ is the complex wavenumber of the mode in air or conductor
  \item $k_0=2\pi f \sqrt{\epsilon_0 \mu_0}=2\pi f/c$ is the free-space wavenumber and $c$ is the speed of light in vacuum.
  \item $\epsilon_r^{(a)} \simeq 1$ is the relative permittivity of air.
  \item $\epsilon_r^{(c)}=\epsilon_0-j\frac{\sigma_c}{2\pi f \epsilon_0}$ is the relative permittivity of the conductor with conductivity $\sigma_c$.
  \item $\mu_r^{(c)}$ and $\mu_r^{(a)}$ are the relative permeabilities of conductor and air, and they are both $\simeq1$.
\end{itemize}
Equation \eqref{eq.SW.TMeqIMPLICIT} can now be solved numerically, for example via fixed point iteration. A good starting point to initialize the iteration $\lambda_a^{(n+1)}=f(\lambda_a^{(n)})$  can be set by recognizing that, for a metal conductor with very high conductivity, one must have $\lambda_a \simeq 0 \Rightarrow h^2=k_a^2=k_0^2 \epsilon_r^{(a)}\mu_r^{(a)} \simeq k_0^2$. Thus, choosing  a value of $h^2=0.99k_0^2$, we can finally set $\lambda_a^{(0)} \simeq \sqrt{k_0^2-0.99k_0^2} = 0.1 k_0 $. This choice allows good convergence over a wide range of frequency.

\subsubsection{Numerical issues}
The ratio of Bessel functions that appears in \eqref{eq.SW.TMeqIMPLICIT} may give rise to numerical issues when the argument $\lambda_c a$ becomes very large, as it happens at THz frequencies and above. To solve this problem we employed the approximation suggested by Orfanidis \cite{orfanidis2002electromagnetic}, based on the asymptotic expansion of Bessel functions \cite{abramowitz1970handbook}. A recent interesting alternative is to use the Lambert-W function \cite{Jaisson2017Lambert}.
\insertonefig {0}{8.8cm}{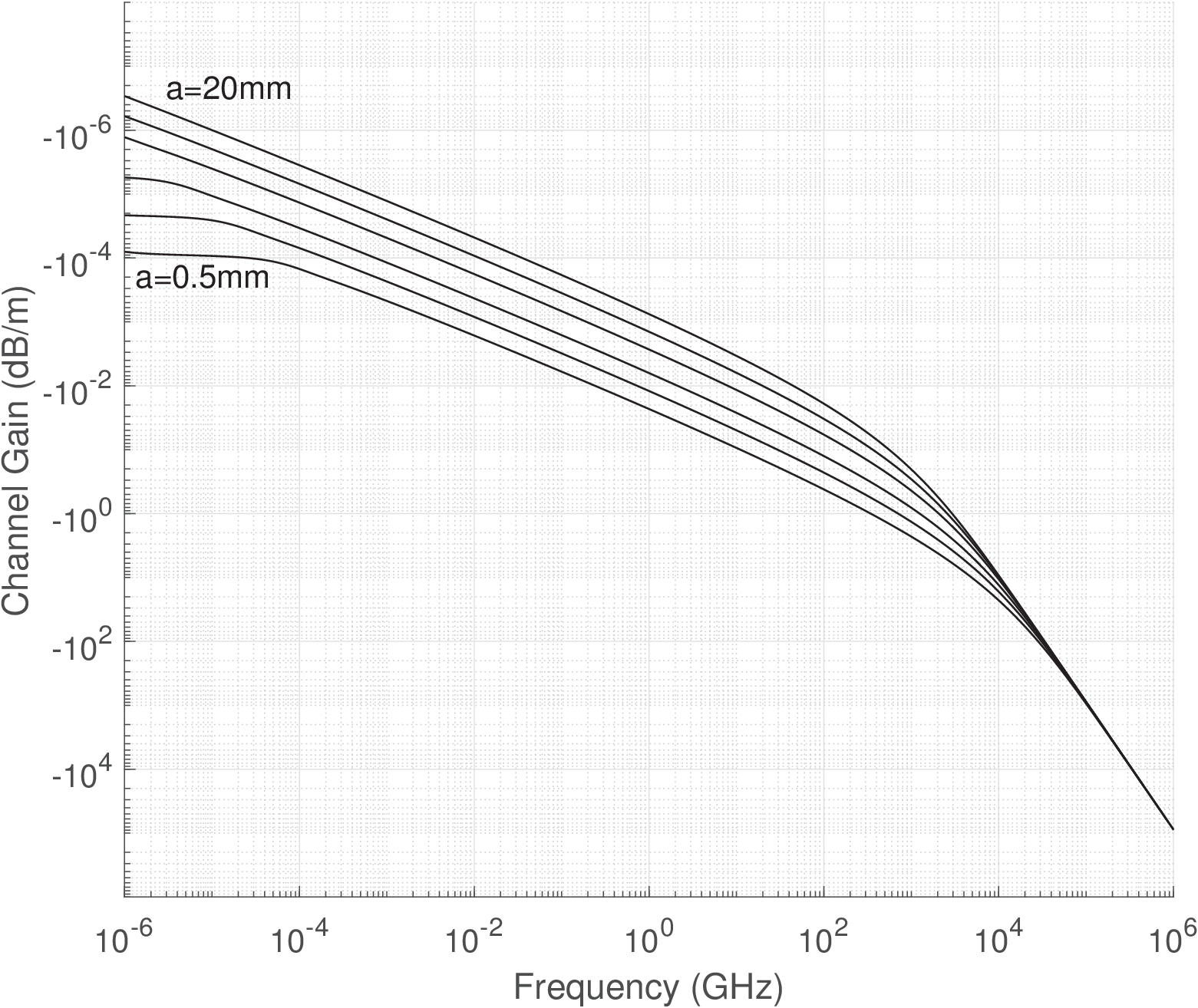}{\label{fig.SWcopperWideFreq}}
{Channel gain of bare copper open waveguide  of radius $a=0.5,~1,~2,~5,~10,~20$~mm.}
\subsection{Axial propagation constant and lateral field extension}
Once $\lambda_a$ is found, other quantities of interest to  communications engineers can be derived. The axial propagation constant $h$ and the radial propagation constant inside the conductor $\lambda_c$ are easily found as:
\begin{IEEEeqnarray}{rCl}
   h &=& \sqrt{k_0^2 \epsilon_r^{(a)}\mu_r^{(a)}-\lambda^2_a} \label{eq.SW.axialprop} \\
  \lambda_c  &=& \sqrt{k_c^2-h^2} = \sqrt{k_0^2(\epsilon_r^{(c)}\mu_r^{(c)}-\epsilon_r^{(a)}\mu_r^{(a)})+\lambda_a^2}. \label{eq.SW.radialprop}
\end{IEEEeqnarray}
The following quantities can be derived from $h$:
\begin{enumerate}
  \item Phase velocity: $v_{ph}(f)=\frac{\omega}{\beta}=\frac{2 \pi f}{\Re[h]}$
  \item Group velocity: $v_{gr}(f)=\frac{\delta\omega}{\delta\beta}$
  \item Attenuation in nepers per meters: $\alpha(f)=-\Im[h]$
\end{enumerate}

The lateral field extension can be easily calculated considering the area of radius $r$ within which a certain percentage of power is contained. The fraction of power flow within the annular area of radius $r$ around the wire is \cite{KingWiltse1962mmw}:
\begin{equation}\label{eq.SW.PowerPct}
   \frac{P^{(r)}}{P_T} = 1-\frac{\Im[r \lambda_a H_1^{(1)}(\lambda_a^* r) H_0^{(1)}(\lambda_a r)]}{\Im[a \lambda_a H_1^{(1)}(\lambda_a^* a) H_0^{(1)}(\lambda_a a)]}
\end{equation}
where $P_T$ is the total transmitted power along the axial direction and $P^{(r)}$ is the power within the annular area of radius $r$ from the wire's surface. By setting the value $\nicefrac{P^{(r)}}{P_T}$ to the power fraction of interest, one can find the radius $r$ of the region within which that fraction of power is contained. For example, if we are interested in the area where $90\%$ of power flows, the equation to solve with respect to $r$ is:
\begin{equation}\label{eq.SW.PowerPctsolve}
    \frac{\Im[r \lambda_a H_1^{(1)}(\lambda_a^* r) H_0^{(1)}(\lambda_a r)]}{\Im[a \lambda_a H_1^{(1)}(\lambda_a^* a) H_0^{(1)}(\lambda_a a)]}=0.1.
\end{equation}

In the mmW range of interest to 5G ($30-100$ GHz), we can see that the line attenuation is a fraction of a dB/m with a lateral field extension in the range of $6-60$ cm depending on cable thickness. The group velocity $v_{gr}(f)$ normalized to the speed of light in vacuum has also been calculated and it is shown in Figure \ref{fig.GroupVelocity}.

The propagation in the bare metal open waveguide is very close to the speed of light in vacuum for a wide frequency range. This allows for reduced latency compared to conventional transmission lines like twisted-pairs and coax whose propagation speed is around $2/3$ the speed of light in vacuum.

Figures \ref{fig.SWcopperWideFreq}-\ref{fig.GroupVelocity} are an example of the capability of the approach followed here, where attenuation in dB/m and lateral field extension can be found for a very wide range of frequencies.

\insertonefig {0}{8.8cm}{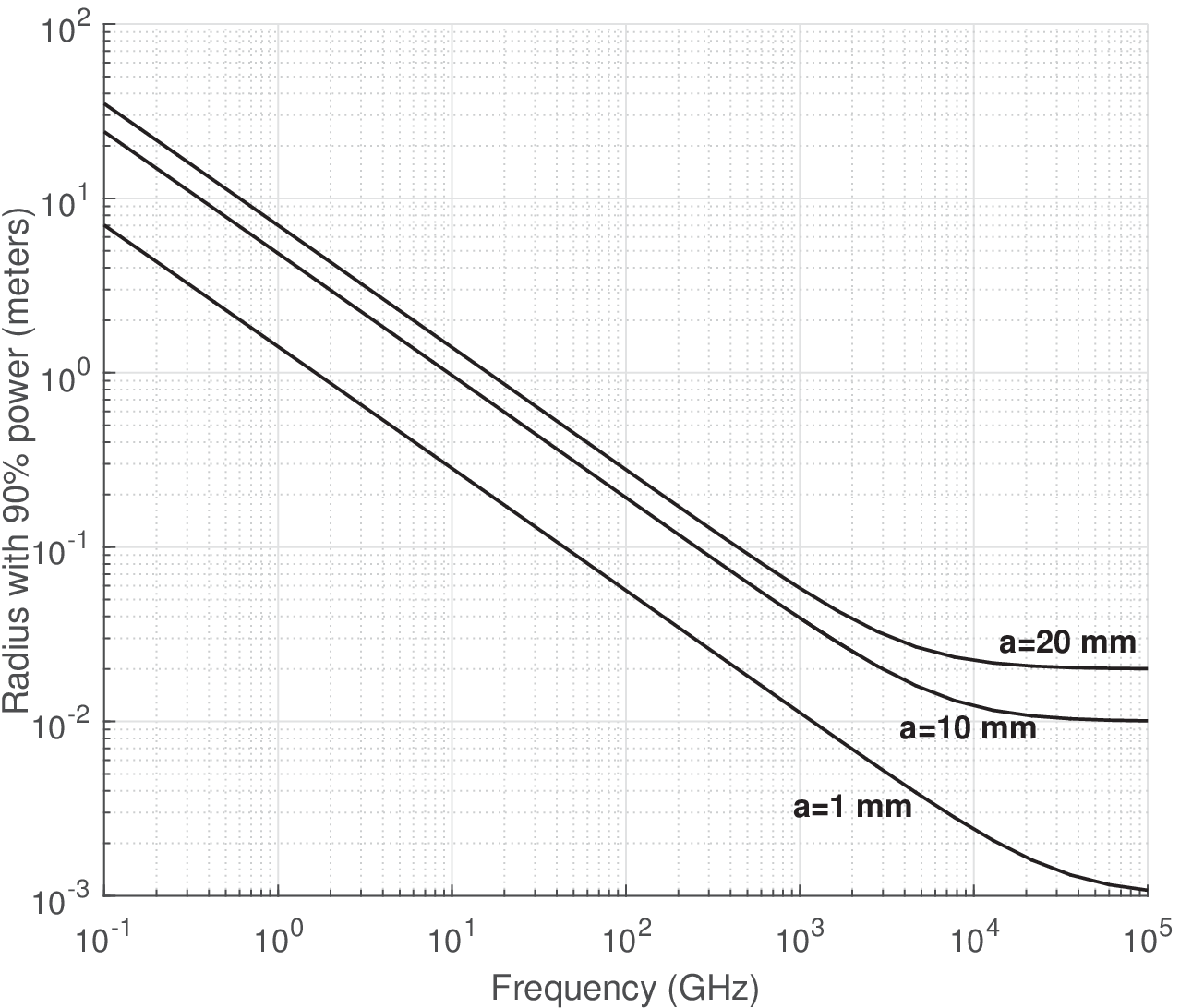}{\label{fig.LateralField90}}
{Radius of area surrounding a copper conductor of radius $a$ within which 90\% of the power is propagated.}
\insertonefig {0}{8.8cm}{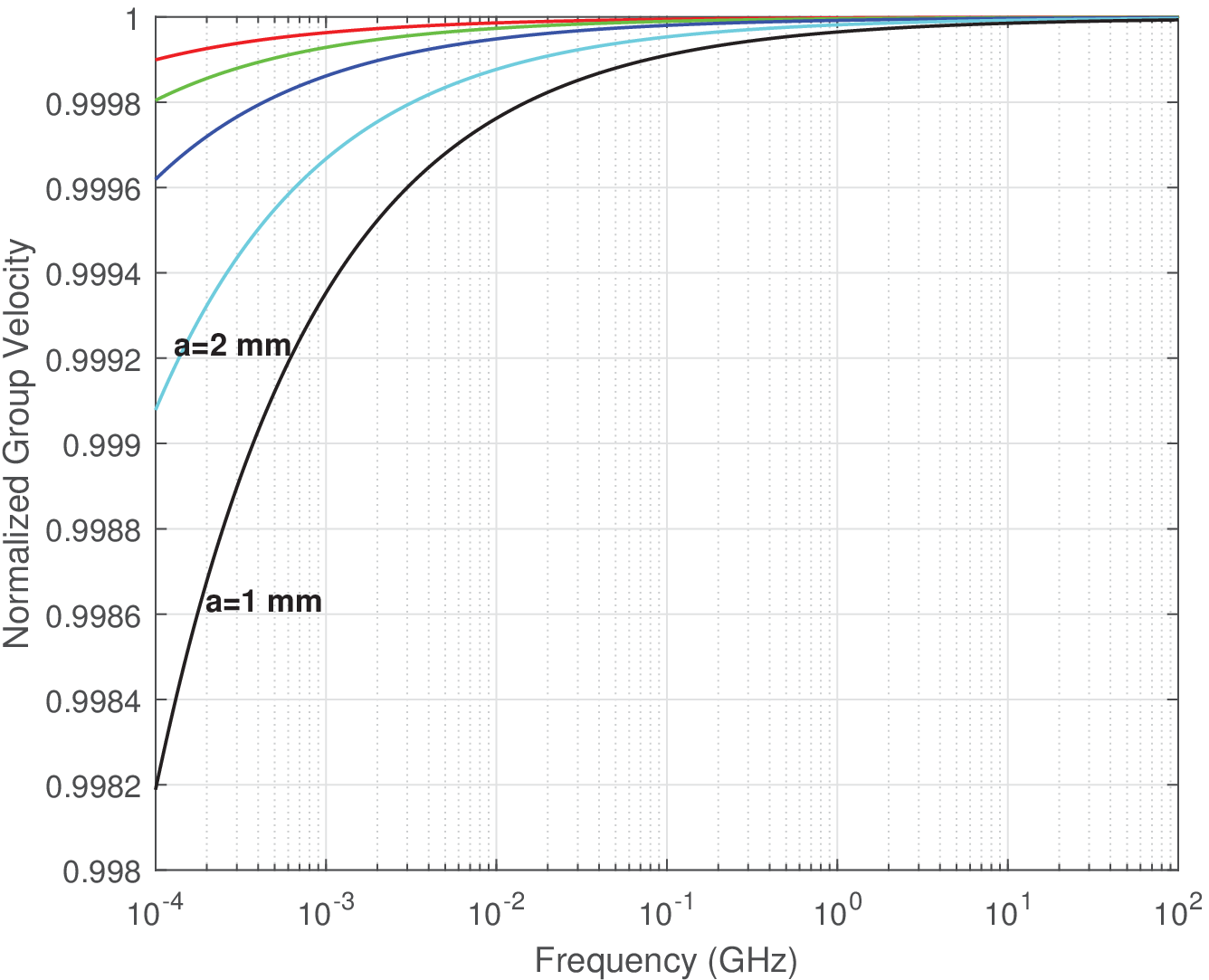}{\label{fig.GroupVelocity}}
{Group velocity normalized to the speed of light in vacuum versus frequency, for radius $a=1,~2,~5,~10,~20$~mm.}

\section{Channel Characteristics}
An analytical expression for the transfer function will be given for the first time in the next sub-section. The ISI will be quantified in terms of Root-Mean-Square Delay-Spread (RMS-DS), which is the square root of the second central moment of the power-delay profile of the channel. We will here consider the band $1-100$~GHz and also compute the channel Shannon capacity. As no information appears to be available on the noise environment of MV lines at GHz frequencies, for capacity calculations it will be assumed that the Power Spectral Density (PSD) of the noise is equal to $-120$~dBm/Hz.

\subsection{Modeling of the Transfer Function} \label{sec.sub.TransferFunction}
Figure \ref{fig.SWcopperMMW} shows the channel gain reported in Figure \ref{fig.SWcopperWideFreq} in the frequency band $1-100$~GHz. From the Figure, we can appreciate that the channel gain in dB/m and the frequency in Hz are very close to be linearly dependent when plotted in a log-log scale. The transfer function in dB of a link of length $d$ can then be expressed as:
\begin{IEEEeqnarray}{rCl}\label{eq.SW.HfdB}
  H^{(dB)}(f) &=& 20\log_{10}(e) \Im [h] d   \\
              &=& - \frac{1}{10^{q(a)}} f^{-m(a)} d
\end{IEEEeqnarray}
where $h$ is the axial propagation constant defined in Sect. \ref{sec.ProblemForm}, $f$ is expressed in Hz, and $d$ in meters. The parameters $m(a)<0$ and $q(a)>0$ depend on the conductor radius $a$ and can be calculated from:
\begin{equation}\label{SW.Hfline}
  -\log(-H^{(dB)}(f)) = m(a) \log(f)+q(a).
%  -\log(k\Im[h]) = m(a) \log(f)+q(a)
\end{equation}
For example, in the frequency interval $1-100$~GHz and for $a=0.5$~mm, we have $m=-0.66$ and $q=7.66$.
\insertonefig {0}{8.8cm}{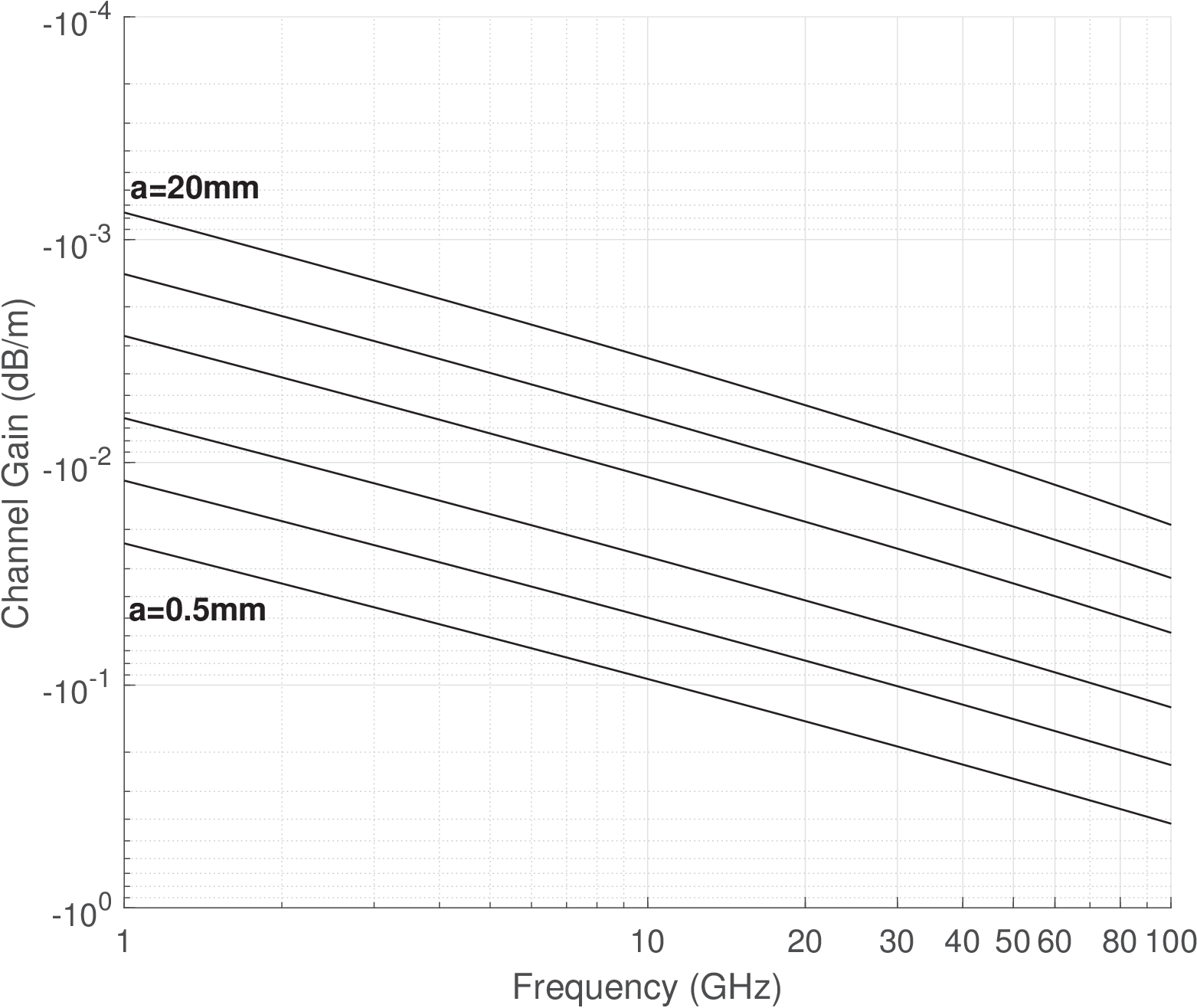}{\label{fig.SWcopperMMW}}
{Same as Figure \ref{fig.SWcopperWideFreq}, for the band $1-100$~GHz.}

\subsection{The Impulse Response} \label{sec.sub.ImpulseResponse}
The impulse response is very similar to a Dirac for short distances and tends to attenuate and broaden as the signal travels along the line. A plot of impulse responses at distances ranging from $50$~m to $500$~m is shown in Figure \ref{fig.SWht}. The amplitude is visibly decreasing at longer distances, but it is not possible to appreciate the pulse broadening on the $\mu s$ scale.

In Figure \ref{fig.SWrmsDSatt}, we show the linear relationship between the logarithm of the RMS-DS and the average channel power gain $\overline{G}= 10 \log (\frac{1}{N} \sum_{i=0}^{N-1} |H(f_i)|^2)$. The same linear relationship was unveiled in \cite{Galli2011wirelinemodel} for a variety of wired and wireless links.
\insertonefig {0}{8.8cm}{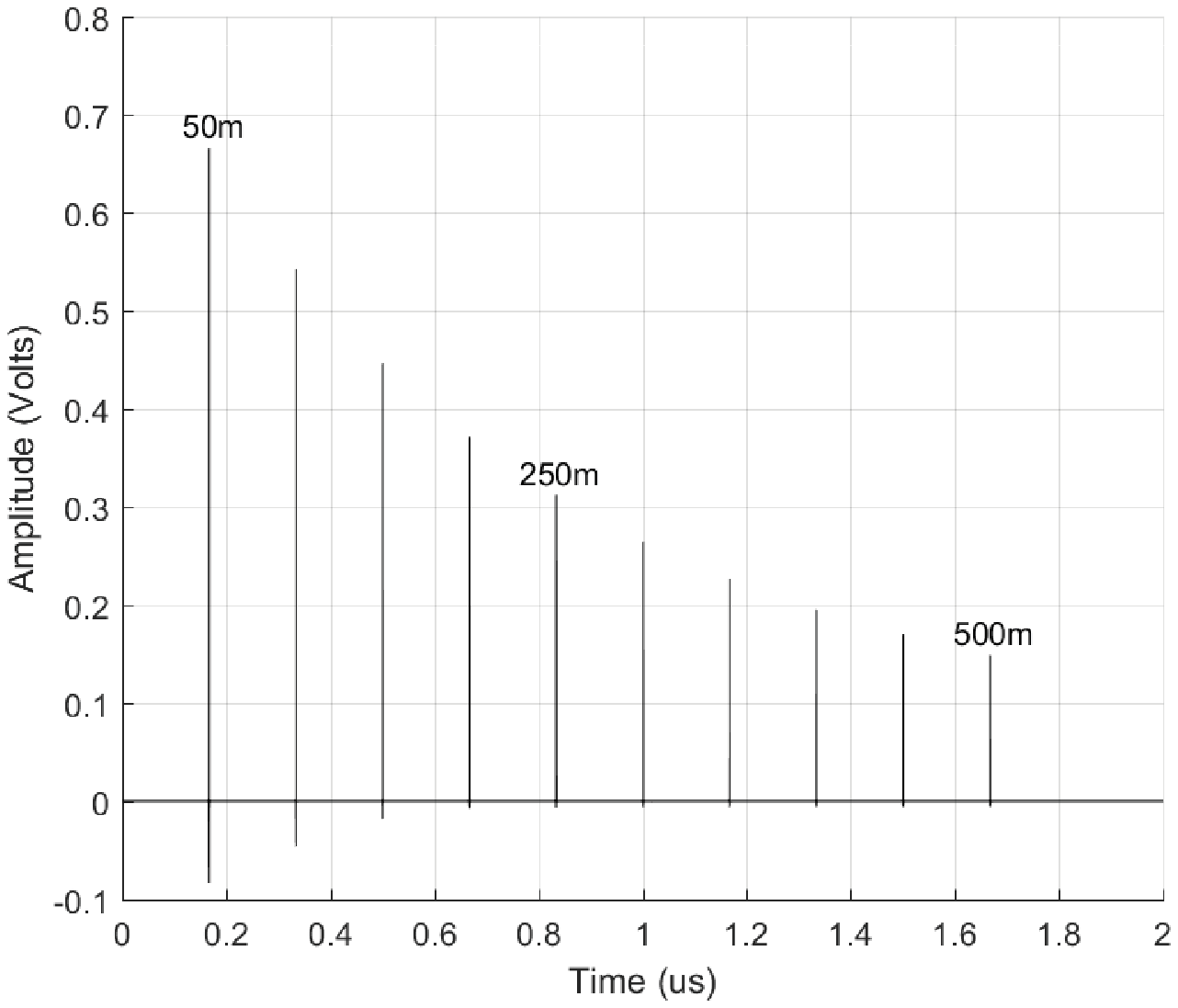}{\label{fig.SWht}}
{Impulse responses for a bare copper wire of radius $a=0.5$~mm after traveling a distance of $50,~100,~150,~\ldots,~500$~m.}
\insertonefig {0}{8.8cm}{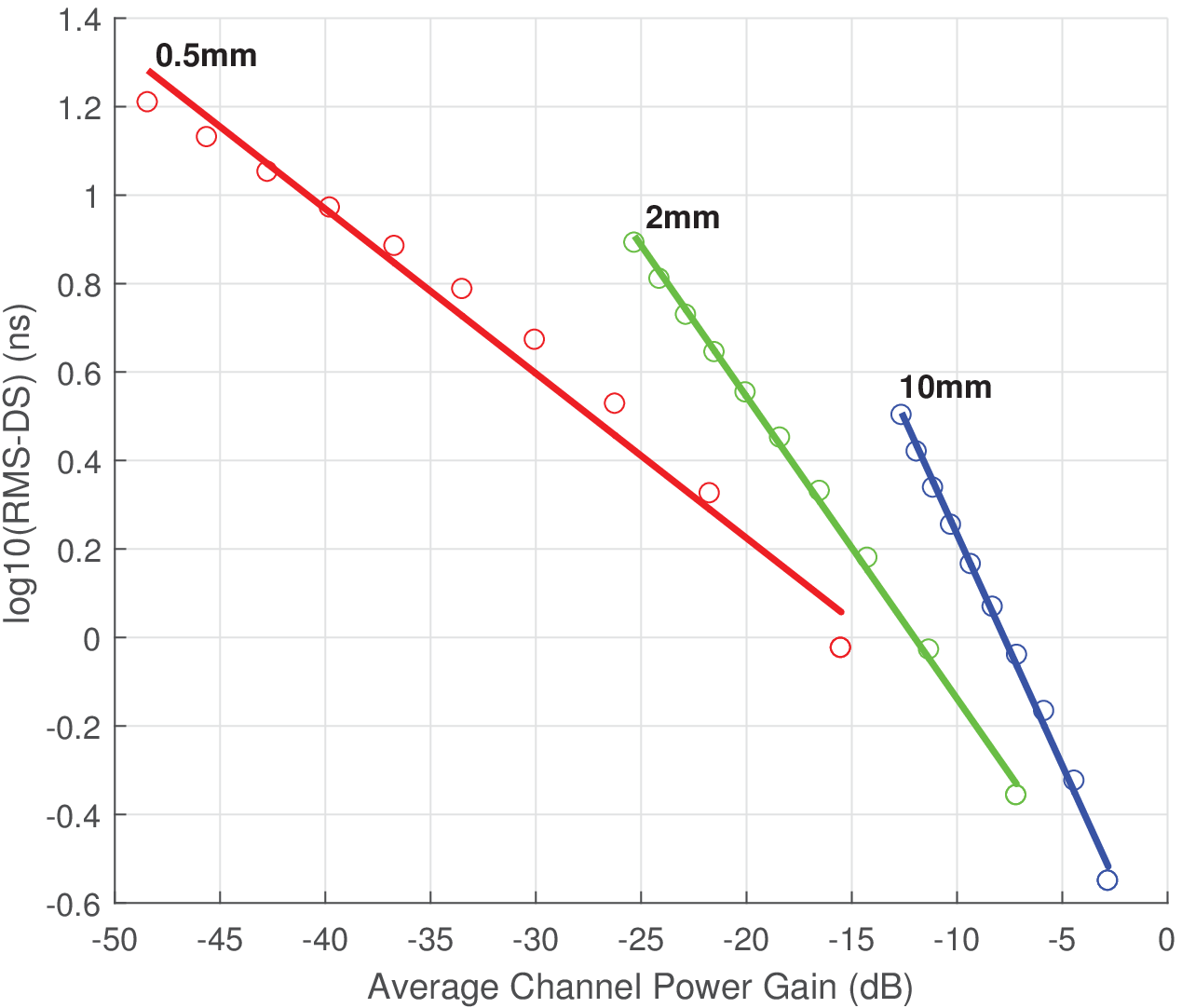}{\label{fig.SWrmsDSatt}}
{Scatter plot of the logarithm of RMS-DS versus $\overline{G}_{dB}$, for copper waveguides of different radii and distances between $50$~m and $500$~m.}

\subsection{Statistical Analysis}
We have verified that both the average power gain in dB and the RMS-DS are lognormally distributed. Both the Anderson-Darling and the Shapiro-Wilks tests confirmed lognormality at the $5\%$ significance level. Similar results for a variety of transmission lines were reported in \cite{Galli2009lognormal,Galli2011wirelinemodel}.

\subsection{Channel Capacity}
In a first set of simulations, we have calculated the achievable channel capacity, which can be expressed as:
\begin{equation}\label{eq:capacity}
    C = W \sum_{k \in \cal K} log_2 \left( 1+\frac{\textit{SNR}_k}{\Gamma} \right)~~~~~bits/sec
\end{equation}
where $k$ is the sub-channel index, $\cal K$ is set of sub-channels carrying information, $W$ is the utilized frequency band, and $\Gamma$ represents an SNR gap factor from Shannon capacity that accounts for the deployment of practical modulation and coding schemes \cite{Book:StaCioSil1999}. This gap measures the efficiency of the transmission scheme with respect to the best possible performance in Additive White Gaussian Noise (AWGN) and can be approximated in dB as follows: $\Gamma_{dB} \approx 9.8 + \gamma_m - \gamma_c$, where $\gamma_c$ is the  coding gain in dB ensured by the employed line code and $\gamma_m$ is the desired system margin in dB.

For the simulations, we have used: $\gamma_m=6$~dB, $\gamma_c=8.8$~dB, a band of $1-100$~GHz, a white noise power density of $N_o=-120$~dBm/Hz. A practical spectral efficiency cap of $12$~bits/sec/Hz was also imposed.

Figure \ref{fig.SWcapacity} shows the enormous carrying capacity of a bare copper wire which can achieve between a few hundred Gbps to $1.2$~Tbps at reasonable distances. On a $10$~mm radius cable and at $100$~m, a data rate of $1$~Tbps could be reached with just $1$W of total transmit power.
\insertonefig {0}{8.8cm}{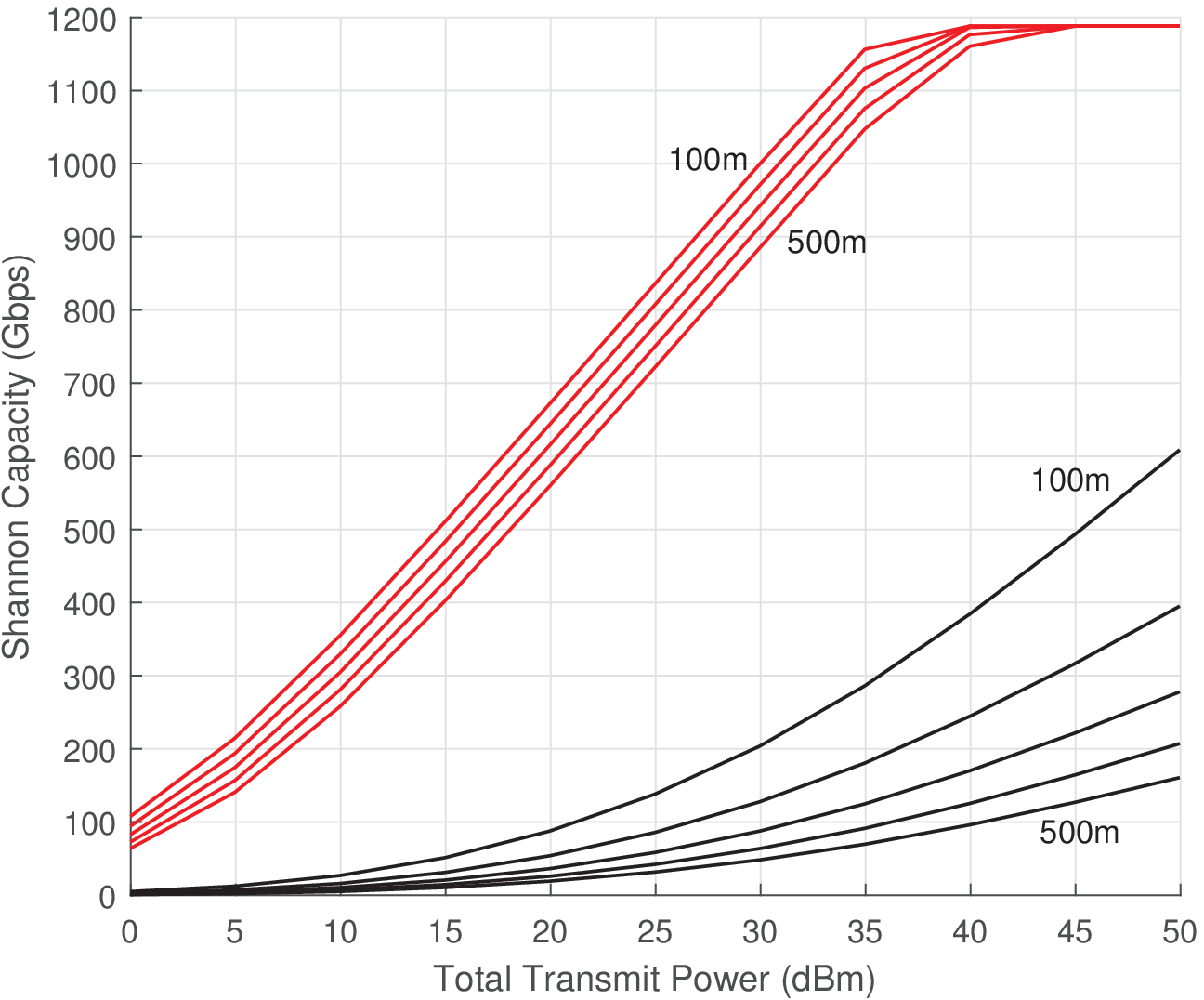}{\label{fig.SWcapacity}}
{Capacity of a bare copper wires of $0.5$~mm (black) and $10$~mm (red) radius at various distances in the range $100-500$~m.}

\section{Effect of Bends and Discontinuities}
Discontinuities along the line (splices, water, nearby objects) disturb the power flow around the wire converting the power of the principal mode into other (usually lossy) modes or they can also cause additional power loss by sky radiation. Furthermore, the important issue of how much degradation is caused by moist and rain is not yet well known, although the little data available shows substantial degradation at cmW frequencies \cite{Kiely1953rain, Chavance1953rain}. These aspects are not included for now in the presented analysis and are left for further study.

The degradation due to rain is not specific to bare metal waveguides, and it is also present in conventional wireless communications at mmW. To solve this problem, one can add a Free Space Optical (FSO) link for diversity to provide a fall back to either link when it rains or there is fog \cite{Link:Collinear}.

\section{Conclusions}
This paper addressed the problem of modeling a single conductor bare metal open waveguide. Numerical techniques have been used to avoid frequency-dependent approximations, thus allowing solving the wave characteristic equations over a very wide range of frequencies (Hz to PHz) and arriving at an analytical expression for the channel transfer function. The best frequency range is in the 5G mmW range and above, where low attenuation and low lateral field extension are ensured together with very high channel capacity.

This kind of open waveguide can be an excellent solution in support of backhauling and broadband access applications in suburban and rural scenarios, where there is an abundance of overhead bare metal power lines.

% If using Bibtex
\bibliographystyle{IEEEtran}
\bibliography{IEEEabrv,All_Papers,PLC_papers}

\end{document}

%% file: MyMacros.tex
\newfont{\bbb}{msbm10 scaled 700}

\newfont{\set}{msbm10 scaled 1100}
\newcommand{\bc}{\mathbf{c}}

\newcommand{\br}{\mathbf{r}}

\newcommand{\epsilonv}{\hbox{\boldmath$\epsilon$}}

\newcommand{\muv}{\hbox{\boldmath$\mu$}}

\newcommand{\sigmav}{\hbox{\boldmath$\sigma$}}

\renewcommand{\Re}{{\rm Re}}
\renewcommand{\Im}{{\rm Im}}

\definecolor{myRED}{RGB}{255, 0, 0}
\definecolor{myBLUE}{RGB}{0,0,255}

%% file: Galli__Final_ICC18__-_Bare_Metal_Open_Waveguide_vfinal_-_ArXiv.bbl
\begin{thebibliography}{10}
\providecommand{\url}[1]{#1}
\csname url@samestyle\endcsname
\providecommand{\newblock}{\relax}
\providecommand{\bibinfo}[2]{#2}
\providecommand{\BIBentrySTDinterwordspacing}{\spaceskip=0pt\relax}
\providecommand{\BIBentryALTinterwordstretchfactor}{4}
\providecommand{\BIBentryALTinterwordspacing}{\spaceskip=\fontdimen2\font plus
\BIBentryALTinterwordstretchfactor\fontdimen3\font minus
  \fontdimen4\font\relax}
\providecommand{\BIBforeignlanguage}[2]{{%
\expandafter\ifx\csname l@#1\endcsname\relax
\typeout{** WARNING: IEEEtran.bst: No hyphenation pattern has been}%
\typeout{** loaded for the language `#1'. Using the pattern for}%
\typeout{** the default language instead.}%
\else
\language=\csname l@#1\endcsname
\fi
#2}}
\providecommand{\BIBdecl}{\relax}
\BIBdecl

\bibitem{Andrews2014What5G}
J.~G. Andrews, S.~Buzzi, W.~Choi, S.~V. Hanly, A.~Lozano, A.~C. Soong, and
  J.~C. Zhang, ``What will {5G} be?'' \emph{{IEEE} J. Sel. Areas Commun.},
  vol.~32, no.~6, pp. 1065--1082, Jun. 2014.

\bibitem{Jaber2016-5Gbackhaul}
M.~Jaber, M.~A. Imran, R.~Tafazolli, and A.~Tukmanov, ``{5G Backhaul Challenges
  and Emerging Research Directions: A Survey},'' \emph{{IEEE Access}}, vol.~4,
  pp. 1743--1766, Apr. 2016.

\bibitem{TR:Corridor:surface2009}
\BIBentryALTinterwordspacing
G.~Elmore, ``{Introduction to the Propagating Wave on a Single Conductor},''
  {Corridor Systems, Inc.}, {White Paper}, Jul. 2009. [Online]. Available:
  \url{http://www.corridor.biz/FullArticle.pdf}
\BIBentrySTDinterwordspacing

\bibitem{Link:AirGig-2016}
\BIBentryALTinterwordspacing
{AT\&T Press Release (Sep. 2016)}. ``\textit{AT\&T Labs' Project AirGig Nears
  First Field Trials for Ultra-Fast Wireless Broadband Over Power Lines}''.
  [Online]. Available:
  \url{http://about.att.com/newsroom/att_to_test_delivering_multi_gigabit_wireless_internet_speeds_using_power_lines.html}
\BIBentrySTDinterwordspacing

\bibitem{GalliScagWang2011ProcIEEE}
S.~Galli, A.~Scaglione, and Z.~Wang, ``{For the Grid and Through the Grid: The
  Role of Power Line Communications in the Smart Grid},'' \emph{Proc. {IEEE}},
  vol.~99, no.~6, pp. 998--1027, Jun. 2011.

\bibitem{Sommerfeld1899}
A.~Sommerfeld, ``{Ueber die Fortpflanzung elektrodynamischer Wellen l\"angs
  eines Drahtes},'' \emph{Ann. Physik Chem}, vol. 303, no.~2, pp. 233--290,
  1899.

\bibitem{Goubau1950SWapplications}
G.~Goubau, ``Surface waves and their application to transmission lines,''
  \emph{Journal of Applied Physics}, vol.~21, no.~11, pp. 1119--1128, 1950.

\bibitem{Goubau1951SC-SWTL}
------, ``Single-conductor surface-wave transmission lines,''
  \emph{{Proceedings of the IRE}}, vol.~39, no.~6, pp. 619--624, Jun. 1951.

\bibitem{Dyott1952launching}
R.~Dyott, ``The launching of electromagnetic waves on a cylindrical
  conductor,'' \emph{Proc. IEE-Part III: Radio and Commun. Engin.}, vol.~99,
  no.~62, pp. 408--413, Oct. 1952.

\bibitem{Goubau1956OpenWires}
G.~Goubau, ``Open wire lines,'' \emph{{IRE Trans. Microw. Theory Tech.}},
  vol.~4, no.~4, pp. 197--200, Oct. 1956.

\bibitem{Brown1959Annular}
J.~Brown and H.~S. Stachera, ``Annular-slot launchers for single-conductor
  transmission lines,'' \emph{{Proc. of the IEE - Part B: Electr. and Commun.
  Engin.}}, vol. 106, no.~13, pp. 143--145, Jan. 1959.

\bibitem{Elmore2012qex}
\BIBentryALTinterwordspacing
G.~Elmore and J.~Watrous, ``A surface wave transmission line,'' \emph{QEX
  Magazine}, pp. 3--9, May/June 2012. [Online]. Available:
  \url{http://www.sonic.net/~n6gn/ELMORE.pdf}
\BIBentrySTDinterwordspacing

\bibitem{Wang2004nature}
K.~Wang and D.~M. Mittleman, ``Metal wires for terahertz wave guiding,''
  \emph{Nature}, vol. 432, no. 7015, pp. 376--379, Nov. 2004.

\bibitem{TR:Corridor:DAS2011}
\BIBentryALTinterwordspacing
G.~Elmore, ``{A New Approach to Outdoor DAS Network Physical Layer Using E-Line
  Technology},'' {Corridor Systems}, Tech. Rep., Mar. 2011. [Online].
  Available: \url{http://www.corridorsystems.com/pdf/DASWhitePaper.pdf}
\BIBentrySTDinterwordspacing

\bibitem{TR:ASSIA:TDSL2017}
\BIBentryALTinterwordspacing
J.~Cioffi, ``{Terabit DSL (TDSL)},'' {ASSIA, Inc.}, Paris, France, {Keynote at
  the G.fast Summit}, May 10, 2017. [Online]. Available:
  \url{http://www.assia-inc.com/wp-content/uploads/2017/05/TDSL-presentation.pdf}
\BIBentrySTDinterwordspacing

\bibitem{SarkarPalma2017}
T.~K. Sarkar, M.~N. Abdallah, M.~Salazar-Palma, and W.~M. Dyab, ``{Surface
  Plasmons-Polaritons, Surface Waves, and Zenneck Waves: Clarification of the
  terms and a description of the concepts and their evolution},'' \emph{{IEEE}
  Antennas Propag. Mag.}, vol.~59, no.~3, pp. 77--93, Jun. 2017.

\bibitem{Stratton1941book}
J.~A. Stratton, \emph{Electromagnetic theory}.\hskip 1em plus 0.5em minus
  0.4em\relax New York and London: McGraw-Hill Book Company, 1941.

\bibitem{KingWiltse1962mmw}
M.~King and J.~Wiltse, ``Surface-wave propagation on coated or uncoated metal
  wires at millimeter wavelengths,'' \emph{{IRE Trans. Antennas Propag.}},
  vol.~10, no.~3, pp. 246--254, May 1962.

\bibitem{orfanidis2002electromagnetic}
S.~J. Orfanidis, \emph{Electromagnetic waves and antennas}.\hskip 1em plus
  0.5em minus 0.4em\relax New Brunswick, NJ: Rutgers University New Brunswick,
  NJ, 2002.

\bibitem{abramowitz1970handbook}
M.~Abramowitz and I.~Stegun, \emph{{Handbook of Mathematical Functions}}.\hskip
  1em plus 0.5em minus 0.4em\relax New York: Dover, 1972.

\bibitem{Jaisson2017Lambert}
D.~Jaisson, ``Wide-range closed formulas for the wavenumber of {Sommerfeld}'s
  wire,'' \emph{{IEEE} Trans. Microw. Theory Techn.}, vol.~65, no.~3, pp.
  693--698, Mar. 2017.

\bibitem{Galli2011wirelinemodel}
S.~Galli, ``A novel approach to the statistical modeling of wireline
  channels,'' \emph{{IEEE} Trans. Commun.}, vol.~59, no.~5, pp. 1332--1345, May
  2011.

\bibitem{Galli2009lognormal}
------, ``A simplified model for the indoor power line channel,'' in \emph{IEEE
  Intl. Symp. on Power Line Commun.\ and Its Appl. (ISPLC)}, Dresden, Germany,
  Mar. 29 -- Apr. 1, 2009.

\bibitem{Book:StaCioSil1999}
T.~Starr, J.~Cioffi, and P.~Silverman, \emph{Understanding Digital Subscriber
  Line Technology}.\hskip 1em plus 0.5em minus 0.4em\relax Prentice--Hall,
  1999.

\bibitem{Kiely1953rain}
D.~Kiely, ``{Experiments with single-wire transmission lines at 3-cm
  wavelength},'' \emph{{Journal of the British Institution of Radio
  Engineers}}, vol.~13, no.~4, pp. 194--199, Apr. 1953.

\bibitem{Chavance1953rain}
P.~Chavance and B.~Chiron, ``Une {\'e}tude exp{\'e}rimentale de transmission
  d'ondes centim{\'e}triques sur guides d'ondes filiformes,'' \emph{Annals of
  Telecommunications}, vol.~8, no.~11, pp. 367--378, 1953.

\bibitem{Link:Collinear}
\BIBentryALTinterwordspacing
{Collinear Networks}. [Online]. Available: \url{http://www.collinear.com/}
\BIBentrySTDinterwordspacing

\end{thebibliography}
